\documentclass[aps,prb,amsmath,amssymb,superscriptaddress,reprint,
]{revtex4-2}

\usepackage{graphicx}
\usepackage{dcolumn}
\usepackage{bm}
\usepackage{hyperref}
\hypersetup{colorlinks=true,allcolors=blue}

\usepackage{siunitx}
\usepackage{physics}

\usepackage{color}

\usepackage{mathptmx}

\begin{document}

	
	\title{Layer-dependent transport properties in the moir{\'e} of strained\\ homobilayer transition metal dichalcogenides} 

	\author{Chao-Jie Ren}	
	\affiliation{Hebei Provincial Key Laboratory of Photoelectric Control on Surface and Interface, School of Science, Hebei University of Science and Technology, Shijiazhuang 050018, China}
	
	\author{Zhao Gong}
	\affiliation{Hebei Provincial Key Laboratory of Photoelectric Control on Surface and Interface, School of Science, Hebei University of Science and Technology, Shijiazhuang 050018, China}
	
	\author{Hui-Ying Mu}
	\affiliation{Hebei Provincial Key Laboratory of Photoelectric Control on Surface and Interface, School of Science, Hebei University of Science and Technology, Shijiazhuang 050018, China}
	
	\author{Xing-Tao An}
	\email[Correspondence to: ]{ anxt2005@163.com}
	\affiliation{Hebei Provincial Key Laboratory of Photoelectric Control on Surface and Interface, School of Science, Hebei University of Science and Technology, Shijiazhuang 050018, China}
	
	\author{Wang Yao}
	\affiliation{Department of Physics and Center of Theoretical and Computational Physics, University of Hong Kong, Hong Kong, China}
	
	\author{Jian-Jun Liu}
	\affiliation{Physics Department, Shijiazhuang University, Shijiazhuang 050035, China}

	\date{\today}
	
\begin{abstract}
		Bilayer moir{\'e} structures have attracted significant attention recently due to their spatially modulated layer degrees of freedom. However, the layer-dependent transport mechanism in the moir{\'e} structures is still a problem to be explored. Here we investigate the layer-dependent transport properties regulated by the strain, the interlayer bias and the number of moir{\'e} periods in a strained moir{\'e} homobilayer TMDs nanoribbon based on low-energy efficient models. The charge carriers can pass perfectly through the scattering region with the moir{\'e} potential. While, it is noted that the overall transmission coefficient is mainly contributed from either intralayer or interlayer transmissions. The transition of transport mechanism between intralayer and interlayer transmissions can be achieved by adjusting the strain. The intralayer transmissions are suppressed and one of the interlayer transmissions can be selected by a vertical external electric field, which can cause a controllable layer polarization. Moreover, the staggered intralayer and interlayer minigaps are formed as the number of moir{\'e} periods increases in the scattering region due to the overlap of the wave functions in two adjacent moir{\'e} periods. Our finding points to an opportunity to realize layer functionalities by the strain and electric field.
\end{abstract}
	
	\maketitle
	
	\sloppy
	
\section{\label{sec:Inro}Introduction}
	
	The inevitable lattice constant mismatch and/or interlayer misorientation of adjacent two-dimensional crystals leads to the formation of long-period moir{\'e} pattern, which has emerged as a highly viable platform to explore exciting physical phenomena that are absent in the monolayers \cite{x1,x2,x3,x4,x5,x6,x7,x8,x9,x10,x11,x12,x13,x14}. Among various moir{\'e} pattern, graphene moir{\'e} superlattices, where a plethora of novel electronic phenomena are observed including the fractal quantum Hall effect \cite{x15,x16,x17,x18} and the emergent correlation phenomena in flat minibands at the magic angle twisting \cite{x19,x20,x21,x22}, have attracted considerable interest recently. Moir{\'e} pattern formed in homobilayers of transition metal dichalcogenides (TMDs) \cite{x23,x24,x25,x26,x27,x28,x29,x30,x31}, proves to be another exciting venue for exploration of the exciting superlattice physics, especially towards their spatially modulated valley and layer pseudospin internal degrees of freedom \cite{x26,x27,x28}.
	
	Compared with their monolayer and heterobilayer counterparts, an extra characteristic of homobilayer TMDs 
	moir{\'e} structures is the layer degrees of freedom which can be described by a pseudospin. The spatial variation of local stacking configurations in each moir{\'e} unit cell can drive the particles to reside in either layer at different lateral positions, forming a moir{\'e} defined layer pseudospin texture \cite{x26,x27}. Moir{\'e} pseudo-magnetic field emerges in such layer texture from the real space Berry phase in the adiabatic motion of carrier \cite{x26,x29}, which underlies the topological dispersion found in the lowest energy minibands \cite{x27}. Such nontrivial topology, together with the intrinsic ferromagnetism arising from the Coulomb exchange \cite{x32}, make bilayer TMDs moir{\'e} an ideal platform for experimental investigations of the quantum anomalous Hall effects \cite{x33,x34,x35,x36}.
	
	The layer texture is analogous to spatial texture of spins or magnetic domains, which implies that the fundamental concepts and theoretical ideas of spin electronics explored in such context can be transferred to the study of layer-dependent electronic properties in van der Waals materials, leading to the development of the field known as layer pseudospintronics or layertronics \cite{x37}. However, the layer-dependent transport properties in the moir{\'e} homobilayer TMDs still remain an open issue hard to solve since thousands of atoms are contained in a moir{\'e} supercell.
	
	Here we investigate the layer-dependent transport properties regulated by the strain and interlayer bias in a strained moir{\'e} homobilayer TMDs nanoribbon based on low-energy efficient models. When the charge carriers transport through the domain wall of the moir{\'e} potential, the transition between the interlayer and intralayer transports can be adjusted by the strain and Fermi energy. The intensity and direction of the layer pseudospin polarization can be controlled by the vertical electric field which can polarize the carriers in a certain layer. The periodic interlayer coupling creates layer-dependent minigaps and minibands in the moir{\'e} superlattice. It is important to clarify that these minigaps are responsible for elucidating the phenomenon of zero transmission, distinguishable from conventional band gaps. When the Fermi energy is located in the interlayer minigaps, the carriers transport is confined with in each individual layer. On the other hand, when the Fermi energy is in the intralayer energy minigaps, the interlayer transmission occurs, namely the carriers incident through the bottom layer will exit from the top layer, and vice versa. Our finding is an illuminating example on the advantage of exploiting the layer pseudospin degree of freedom in future electronics.
	
	\begin{figure}
		\includegraphics[width=8.5cm]{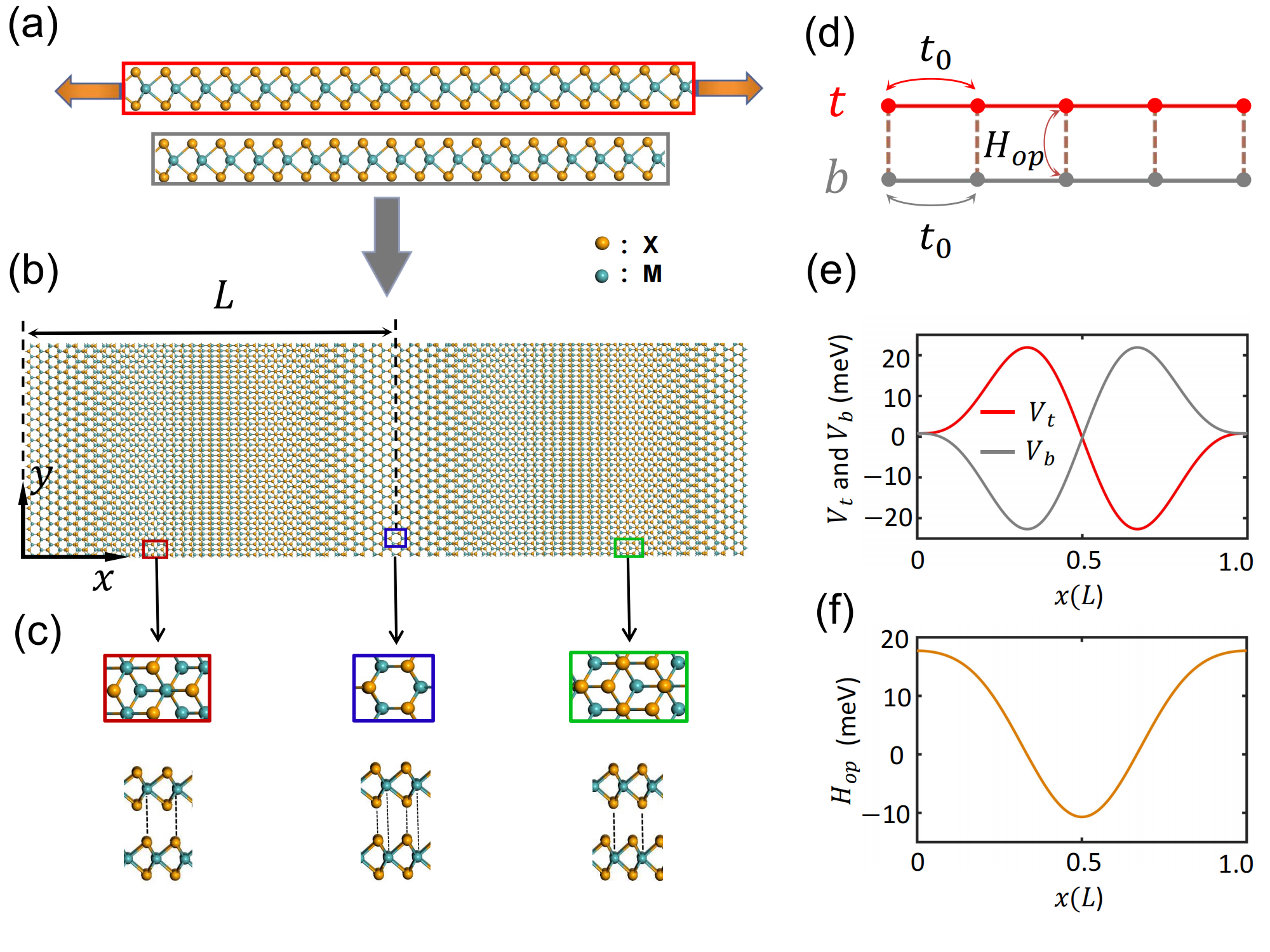}
		\caption{\label{fig:a1}The schematics and potentials of strained moir{\'e} superlattice. (a) Side view sketches of a moir{\'e} pattern of strained homobilayer TMDs. The metal and chalcogen atoms are represented by blue and yellow colors, respectively.  (b) Top view of the moir{\'e} superlattice. The rectangles of different colors show the high symmetry stackings. (c) Zoom in around the atomic structures of different high symmetry points. (d) The discrete lattice model for the effective Hamiltonian. The top and bottom layers are represented by a set of red and grey points, respectively. The hopping term between the nearest points in the same layer is $t_0$ and the coupling in the valence bands with $t/b$ labeling the top/bottom layer is characterized by $H_{op}$. (e) The intralayer moir{\'e} potentials $V_t$ and $V_b$. (f) The interlayer hopping between the valence band edges of the two layers.
		}
	\end{figure}
	
\section{\label{sec:theor}MODEL AND METHODS}
	
	In this work, we delve into the moir{\'e} patterns formed from parallelly stacked homobilayer TMDs, where the armchair and zigzag edges have been assigned as the $\hat{x}$ and $\hat{y}$ axes, respectively. The transport direction is along the $\hat{x}$ axis and the structure is confined along the $\hat{y}$ axis. Considering parallel-aligned stacking of homobilayer TMDs, a uniform strain is applied to the top layer along the $\hat{x}$ direction with the only non-zero component of the strain tensor being $\epsilon_{xx}=\eta$, while no strain is applied to the bottom layer, as shown in Fig.~\ref{fig:a1}(a). In this case, due to lattice constant mismatch, moiré patterns are formed, as illustrated in Fig.~\ref{fig:a1}(b). Given the narrow width of the ribbon, the effect of lattice mismatch along the $\hat{y}$ direction on the observed results can be safely disregarded. The expected strain can be achieved experimentally by employing the method of elongating the substrate \cite{x38}. The moir{\'e} pattern arises from local high-symmetry stacking configurations, namely $R^M_X$, $R^M_M$, and $R^X_M$, as depicted in Fig.~\ref{fig:a1}(b). Here, $R^M_X$($R^X_M$) refers to Bernal stacking, where metal (chalcogen) atoms from the top layer sit atop chalcogen (metal) atoms from the bottom layer, and $R^M_M$ corresponds to aligned parallel stacking, as illustrated in Fig.~\ref{fig:a1}(c).
	
	As the energy gap between the conduction and valence bands of TMDs is considerable, the interband coupling between the two layers can be disregarded, and we will concentrate on the interlayer coupling between the valence band edges from the two layers. The significant spin splitting induced by spin-orbit coupling (SOC) in the valence bands leads to spin-valley locking in the low-energy regime, with spin up/down being tied to valley $-K/K$. In the case of moir{\'e} formed from R-stacking, interlayer coupling occurs between band edges having the same spin and valley indices from the two layers. The effective Hamiltonian governing the valence band of the $K$-valley can be expressed as
	
	\begin{equation}\label{eqi1}
		H_v=-\frac{p^2}{2m}+\mathcal{U}_v=-\frac{p^2}{2m}+\left(\begin{array}{cccc}
			V_t&H_{op}\\
			H^*_{op}&V_b\\
		\end{array}\right) ,
	\end{equation}
	
	\noindent where $-\frac{p^2}{2m}$ describes the quadratic dispersion near the valence band edge, $\mathcal{U}_v$ characterizes the moir{\'e} potential of the valence band coupling, $V_{t/b}$ represents the intralayer moir{\'e} potential, elucidating the band edge shifts in each individual layer due to the charge transfer between layers, and $t/b$ denotes the top/bottom layer\cite{x29}. The $2\times2$ matrix form arises from the internal degrees of freedom of the layer pseudospin, reflecting that electrons can reside in either layer of the homobilayer. A simple discrete lattice model can be used to depict this Hamiltonian, as shown in Fig.~\ref{fig:a1}(d). For reference, the Appendix B provides detailed information on the finite difference method and the discrete lattice model.
	
	The intralayer hopping energy can be modified by strain in moir{\'e} superlattices due to the atomic distance variations within the strained layer. The effect of this phenomenon can be expressed as $A_{\epsilon}={\tau}\frac{\sqrt{3}\hbar\beta}{2a}(\epsilon_{yy}-\epsilon_{xx},-2\epsilon_{yx})$, where $\tau=\pm$ exhibits opposite signs in the two valleys, $\beta=2.4$, and $\epsilon$ represents the strain tensor. As $A_\epsilon$ causes an effective displacement of the $K$ point, the interlayer hopping $H_{op}$ also acquires a strain-dependent phase, i.e., $H_{op}\rightarrow H_{op}e^{-i\frac{\boldsymbol{A_{\epsilon}}\cdot\boldsymbol{r} }{\hbar}}$. The moir{\'e} potential $V_{l=t/b}$ and $H_{op}$ can be modeled as (see Appendix A for details from the two-dimensional moiré potential to the one-dimensional moiré potential)
	
	\begin{equation}\label{eqi2}
		V_l=V_0[2\cos(2\pi\frac{x}{L}+\alpha_l)+\cos(-4\pi\frac{x}{L}+\alpha_l)] ,
	\end{equation}

	\begin{equation}\label{eqi3}
		H_{op}=\{h_0[1+2\cos(2\pi\frac{x}{L})]+h_1[1+2\cos(4\pi\frac{x}{L})]\}e^{-i\frac{\boldsymbol{A_{\epsilon}}\cdot\boldsymbol{r} }{\hbar}},
	\end{equation}
	
	\noindent where $L\approx\frac{\sqrt{3}a}{\eta}$ represents the moir{\'e} period with a monolayer lattice constant $a$, and $\eta$ is the unique non-zero component of the strain tensor, i.e., $\epsilon_{xx}=\eta$\cite{x29}. The values of other parameters are contingent upon the material, and in this work, we have employed the parameters specific to $MoSe_2$: $V_0\approx8.586 meV$, $\alpha\approx\mp0.49\pi$ (corresponding to $l=t$ and $b$, respectively), $h_0\approx7.1 meV$, and $h_1\approx-1.2 meV$\cite{x29}. Fig.~\ref{fig:a1}(e) shows the intralayer moir{\'e} potentials for both the top and bottom layers over a moir{\'e} period, with the extreme values identified at $\frac{L}{3}$ and $\frac{2L}{3}$. Furthermore, Fig.~\ref{fig:a1}(f) demonstrates the moir{\'e} potential relating to the interlayer hopping between the valence band edges of the two layers.

	In the following we will focus on layer-dependent transmission through the region between the two different high symmetry points. Our analysis will specifically examine the effect of the strain tensor component $\eta$ and the Fermi energy $E_f$ on layer-dependent transport. Consider a central scattering region, which has a strong interlayer coupling and is connected to bilayer TMDs source and drain with $R^M_X$ and $R^X_M$ configuration correspondingly, as shown in Fig.~\ref{fig:a2}(a). The energy band diagrams of the source and drain are depicted on both sides of the panel in Fig.~\ref{fig:a2}(a), where the red curves (gray curves) correspond to the energy band of the top (bottom) layer, and the black dashed lines represent the zero-energy reference point. The valence band maximums (VBM) of top and bottom layers in the source are $V_1=21.9 meV$ and $V_2=-22.7 meV$, respectively, whereas the opposite holds true in the drain. The layer-dependent transmission coefficient is calculated with the Hamiltonian in Eq. (1), using a recursive Green's function technique \cite{x39}. For further details, please refer to Appendix C.
	
\section{RESULTS AND DISCUSSION}\label{sec3}

\begin{figure}
	\includegraphics[width=8.5cm]{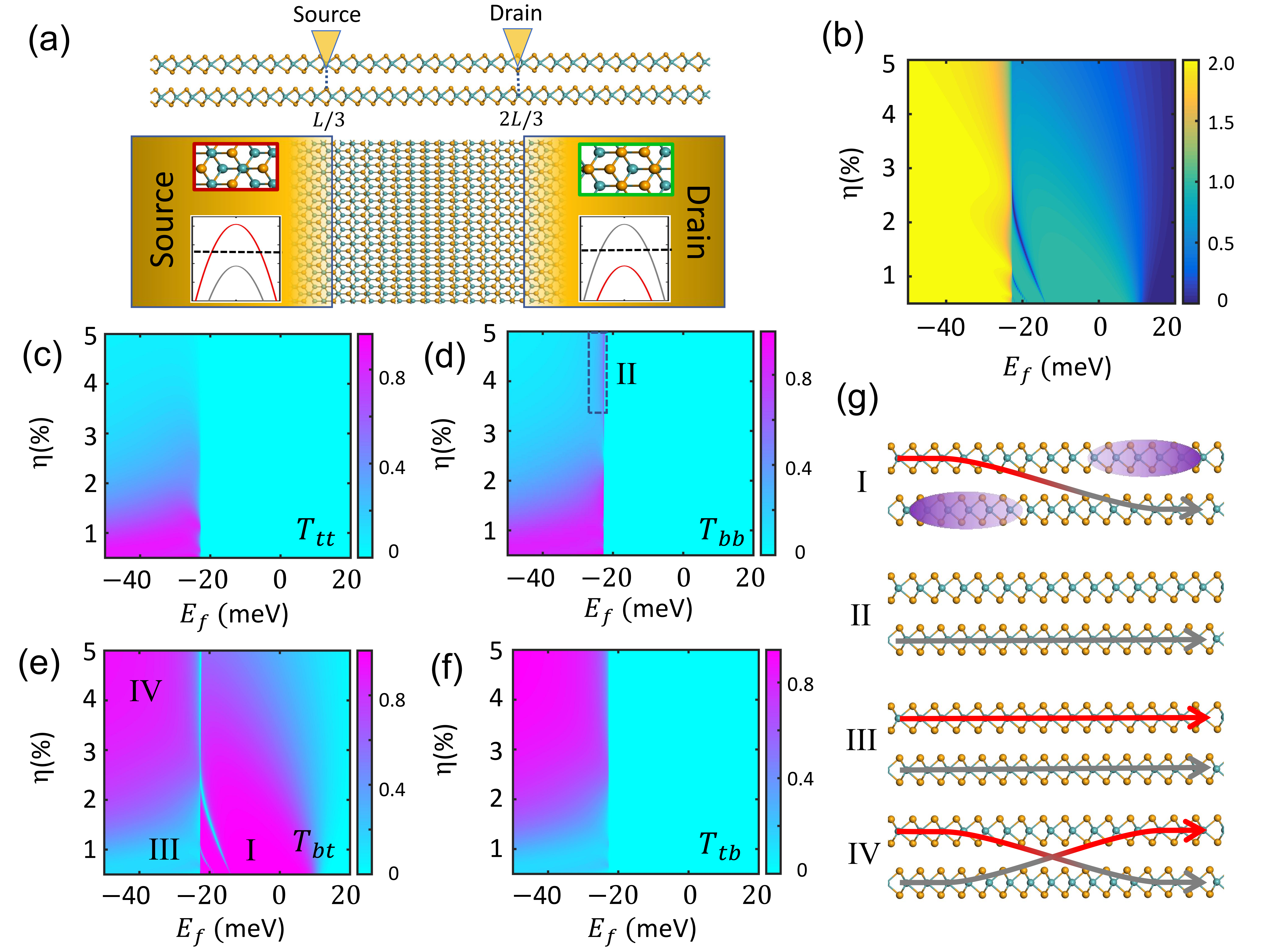}
	\caption{\label{fig:a2}Layer-dependent transmission through the region between the two different high symmetry points. (a) Schematic of a central scattering region with strong interlayer coupling connected to bilayer TMDs source and drain with $R^M_X$ and $R^X_M$ configuration, respectively. (b) The overall transmission coefficient as a function of the Fermi energy $E_f$ and the component $\eta$ of strain tensor. (c-f) The intralayer and interlayer transmission coefficients. (g) Schematics of the intralayer and interlayer transmission in regions I-IV marked in (d) and (e).
	}
\end{figure}

 	Fig.~\ref{fig:a2}(b) shows the overall transmission coefficient $T_{sum}$($T_{sum}=T_{tt}+T_{bb}+T_{bt}+T_{tb}$) as a function of the Fermi energy $E_f$ and the strain tensor component $\eta$, where $T_{ll'}$ is the layer-dependent transmission coefficient for the incident layer $l'$ and out-going layer $l$. Figs.~\ref{fig:a2}(c-f) plot the layer-dependent transmission coefficients as functions of $E_f$ and $\eta$. When the Fermi energy is located between the VBM of the top layer and bottom layer ($V_2<E_f<V_1$), the overall transmission is only contributed by the transmission $T_{bt}$ from the top to the bottom layer, as shown in Figs.~\ref{fig:a2}(b) and \ref{fig:a2}(c-f), because only electronic states of the top (bottom) layer in the source (drain) are involved in transport. As can be seen in Figs.~\ref{fig:a2}(b) and \ref{fig:a2}(e), there are two sets of dips induced by the quantum interference effect in the transmission spectrum in this energy range. The moir{\'e} period is very large when the strain is weak. There are localized states in the top layer near the drain and in the bottom layer near the source. The interference between these localized states and the continuous states (shown in region I in Fig.~\ref{fig:a2}(e)) will form Fano anti-resonance peaks \cite{x40}. As the strain becomes stronger, the size of the scattering region ($\sim\frac{a}{\eta}$) will become smaller, and these localized states will disappear, thus the Fano effect will gradually weaken. Meanwhile, as the scattering region becomes shorter, the interlayer coupling effect becomes weaker, and the transmission becomes smaller.

	When the Fermi energy is around $V_2$, the activation of the new transport mode causes the destructive interference in interlayer transmission \cite{x41}. The intralayer moir{\'e} potentials are steep in the case of strong strain. The difference of the two layers is that the holes in the top layer feel the scattering region as a barrier, while the holes in the bottom layer feel a well. Therefore, in this case, the carriers can only propagate from the bottom layer of the source to the bottom layer of the drain (shown in Figs.~\ref{fig:a2}(d) and \ref{fig:a2}(g) II). Conversely, when the direction of propagation is reversed, carriers can solely propagate from the top layer of the drain to the top layer of the source.

	When the Fermi energy is less than the VBM of the two layers, $E_f<V_2$, we find an almost perfect overall transmission coefficient, shown in Fig.~\ref{fig:a2}(b). An interesting phenomenon that can be observed is that the transition of transport mechanism from intralayer transmission to interlayer transmission appears with increasing the strain. The overall transmission coefficient is mainly contributed from either intralayer or interlayer transmissions although the valence bands of both layers are involved in transport. In this scenario, the intralayer transmission is primarily observed when a weak strain is applied to the strain layer (see region III in Fig.~\ref{fig:a2}(e)). With increasing the strain, interlayer transmission becomes increasingly dominant (see region IV in Fig.~\ref{fig:a2}(e)). The length of scattering region becomes shorter and the quantum confinement effect in the scattering region is strengthened with the increase of the strain. The carrier distribution in the top and bottom layers accumulates towards the center of the scattering region. Since the scattering center is where the interlayer coupling is strongest, the strong strain greatly strengthens the interlayer wave function overlap, then the interlayer transmission plays a dominant role in transport.

\begin{figure}
	\includegraphics[width=8.5cm]{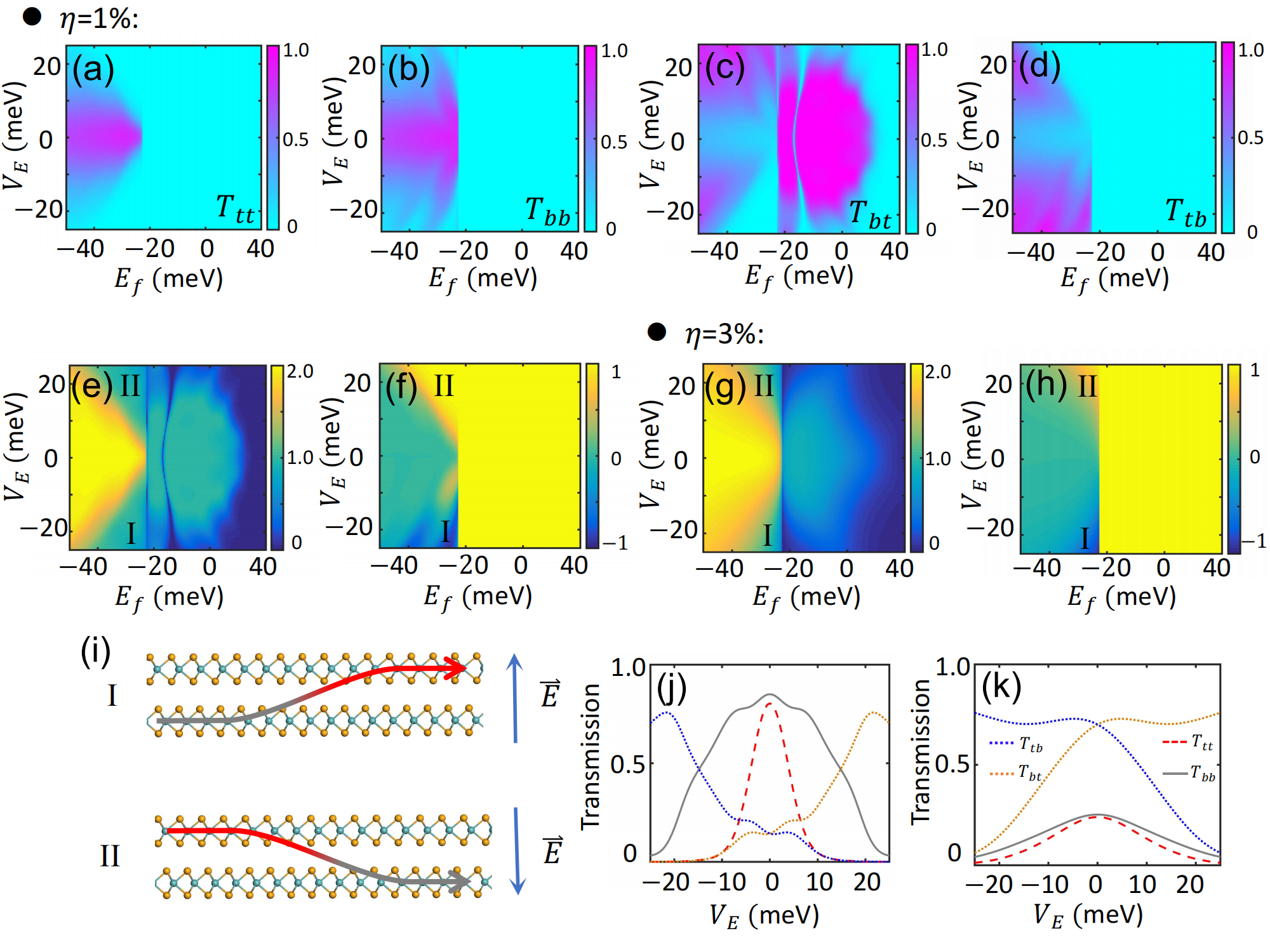}
	\caption{\label{fig:a3}Layer-polarized performance under interlayer bias. (a-d) The intralayer and interlayer transmission coefficients as functions of the Fermi energy $E_f$ and the interlayer bias $V_E$, at $\eta=1\%$. (e) The overall transmission coefficient and (f) The layer polarization $P_l$ as functions of the Fermi energy $E_f$ and the interlayer bias $V_E$ at $\eta=1\%$. (g) The overall transmission coefficient and (h) The layer polarization $P_l$ as functions of the Fermi energy $E_f$ and the interlayer bias $V_E$ at $\eta=3\%$. (i) Schematics of the layer-dependent transmission in regions I-II marked in (f) and (h). (j) and (k) The intralayer and interlayer transmission coefficients as functions of the interlayer bias $V_E$ at $\eta=1\%$ and $\eta=3\%$, respectively. The Fermi energy is set as $E_f=-26meV$.
	}
\end{figure}

Next, we will investigate the influence of the interlayer bias on the layer-dependent transport properties. The two layers are placed at an additional potential of $\pm V_E$ through the application of an interlayer bias with $V_E=\frac{1}{2}eEd$, where $E$ represents the perpendicular electric field and $d$ represents the space between the two layers \cite{x26}. Figs.~\ref{fig:a3}(a-d) illustrate the interlayer and intralayer transmission coefficients as functions of the interlayer bias and Fermi energy when the strain tensor component $\eta=1\%$. When the Fermi energy is located at $E_f<V_2$, the intralayer transmissions are suppressed by an external electric field. Similar to the strain, an applied electric field can also realize the transition of transport mechanism from intralayer transmission to interlayer transmission. However, compared with the strain, the electric field has a more complicated effect on the interlayer transmissions. An electric field in a downward direction enhances the top-to-bottom transmission $T_{bt}$ and suppresses the bottom-to-top transmission $T_{tb}$, while the opposite effect is observed for an electric field in an upward direction, as shown in Fig.~\ref{fig:a3}(i). These effects of the external electric field on the transport properties mainly come from the enhancement of the intralayer potentials and the Coulomb force on the hole induced by the electric field.

Fig.~\ref{fig:a3}(e) shows the corresponding overall transmission coefficient $T_{sum}$, from which it can be seen that there are two triangular regions I and II with low transmission coefficients as $E_f<V_2$ and the external electric field is strong. At these regions, there is only one of the interlayer transmissions $T_{bt}$ or $T_{tb}$, which can give rise to a significant layer polarization. Fig.~\ref{fig:a3}(f) plots the layer polarization $P_l$ as a function of the interlayer bias and the Fermi energy, where $P_l=(T_{bt}+T_{bb}-T_{tt}-T_{tb})/(T_{tt}+T_{bb}+T_{bt}+T_{tb})$. Fig.~\ref{fig:a3}(i) illustrates the transmission schematics of regions I and II in Fig.~\ref{fig:a3}(f). It is evident that the negative layer polarization in region I results from an upward electric field and the transmission solely from $T_{tb}$. Likewise, the positive layer polarization in region II is caused by a downward electric field and the transmission solely from $T_{bt}$.

We also calculate the overall transmission coefficient and layer polarization as functions of the interlayer bias $V_E$ and the Fermi energy $E_f$ for the strain tensor component $\eta=3\%$, which show similar results to those obtained for $\eta=1\%$, as shown in Figs.~\ref{fig:a3}(g-h). Whether it is a strong strain or a weak strain, the intralayer transmissions are suppressed by the interlayer bias (shown in Figs.~\ref{fig:a3}(j) and \ref{fig:a3}(k)). Compared with the results of the weak strain case, the layer polarization regions I and II become smaller in the strong strain case because the interlayer transmission is dominant in the absence of the interlayer bias.

\begin{figure}
	\includegraphics[width=8.5cm]{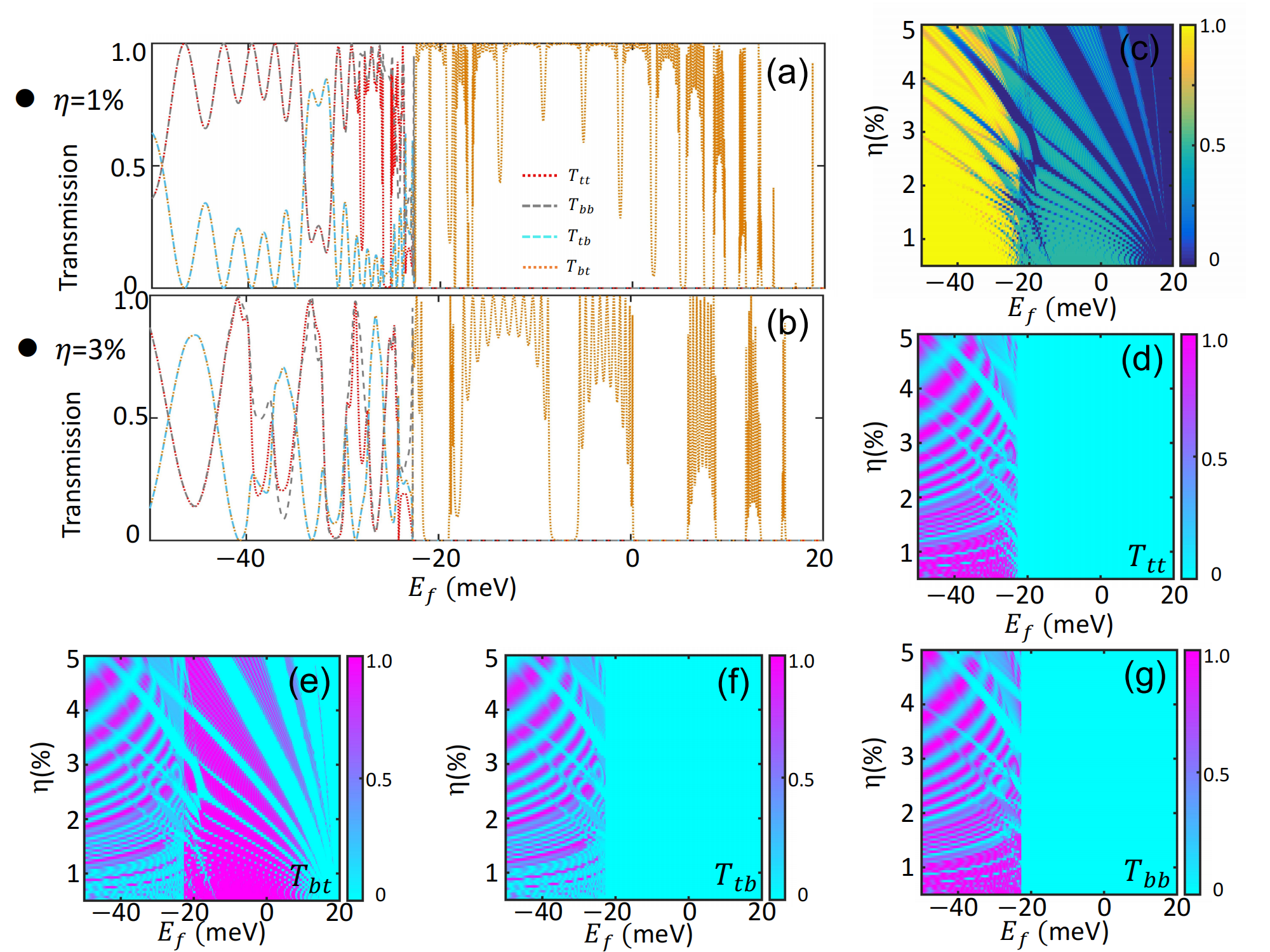}
	\caption{\label{fig:a4}Layer-dependent transmission through strained moir{\'e} superlattice. (a) and (b) The intralayer and interlayer transmission coefficients as functions of the Fermi energy $E_f$ at $\eta=1\%$ and $\eta=3\%$, respectively. (c) The overall transmission coefficient. (d-g) The intralayer and interlayer transmission coefficients as functions of $E_f$ and $\eta$. All calculations here use the number of the moir{\'e} superlattice period $N=10$.
	}
\end{figure}

Finally, we examine how the transport properties of the strained moir{\'e} superlattice are influenced by the number of periods $N$ in the scattering region. Specifically, we focus on $N=10$ and investigate the layer-dependent transport properties. Figs.~\ref{fig:a4}(a) and (b) depict the intralayer and interlayer transmission coefficients as functions of the Fermi energy $E_f$ for $\eta=1\%$ and $\eta=3\%$, respectively. It is evident that both the intralayer and interlayer transmission coefficients exhibit oscillatory behavior when both valence bands of the two layers are involved in transport. For $\eta=1\%$, at the energy corresponding to resonance peaks of the intralayer transmission, the interlayer transmission coefficients vanish and the intralayer transmissions dominate. While, for the strong strain, $\eta=3\%$, the intralayer and interlayer transmissions alternate dominance. When the Fermi energy is located between the two VBM of the two layers, the only form of transmission is interlayer transmission $T_{bt}$, displaying miniband transport phenomena. 

Figs.~\ref{fig:a4}(d-g) plot the intralayer and interlayer transmission coefficients as functions of the Fermi energy $E_f$ and the strain tensor component $\eta$ for $N=10$ in the moir{\'e} superlattice. As the strain increases, the period of the superlattice $L$ becomes smaller and the wave functions of two adjacent moir{\'e} period potential overlap more and more. According to the superlattice transport theory, the moir{\'e} superlattice will have the properties of miniband transport, that is, the system will have layer-dependent minibands and minigaps. The periodic transitions of intralayer and interlayer transport mechanisms are made possible by the staggered intralayer and interlayer gaps, which can be seen obviously in Figs.~\ref{fig:a4}(d-g).

Based on the one-dimensional model, the Fermi energy range employed in this study is $\Delta{E_f}=70$ meV. Through the computation of the energy difference $\Delta{E}=\frac{3\pi^2\hbar^2}{2m^*{L_y}^2} $ between the ground state and the first excited state, it is found that if the $\hat{y}$ direction length, $L_y$, is less than 5.18 nm, there is no introduction of extra sub-bands that could impact the model. When a strain of $1\%$ to $5\%$ is applied to the ribbon, the displacement between the top and bottom layers in the $\hat{y}$ direction ranges from 0.05nm to 0.26nm. Therefore, the effect of lattice mismatch along the $\hat{y}$ direction on the observed results can be safely disregarded. In this case, the obtained results are primarily dependent on the $\hat{x}$-coordinate, as demonstrated by Eqs. (\ref{eqi2}) and (\ref{eqi3}). When the length $L_y$ along the $\hat{y}$ axis exceeds 5.18 nm, the one-dimensional model is no longer applicable due to the impact of additional sub-bands and the effect of lattice mismatch along the $\hat{y}$ direction. Thus, a detailed consideration of the two-dimensional scenario is required, which will be investigated in future work.

\section{CONCLUSIONS} \label{sec4}

In summary, we have investigated the layer-dependent transport properties regulated by the strain, the interlayer bias and the number of moir{\'e} periods in a strained moir{\'e} homobilayer TMDs nanoribbon based on low-energy efficient models. An almost perfect overall transmission coefficient can be found when the valence bands of both layers are involved in transport. However, the overall transmission coefficient is mainly contributed from either intralayer or interlayer transmissions. The transition of transport mechanism from intralayer transmission to interlayer transmission can be achieved by adjusting the strain. The intralayer transmissions can be suppressed by an external electric field. By regulating the direction of the external electric field, one of the interlayer transmissions is selected, which can give rise to a significant layer polarization. Therefore, the direction of layer polarization can be controlled by the electric field. As the number of moir{\'e} periods increases in the scattering region, the staggered intralayer and interlayer minigaps are formed at the strong strain due to the overlap of the wave functions in two adjacent moir{\'e} periods. These results point to an unexpected but exciting opportunity to build layer functionality by the strain and electric field in moir{\'e} structures.

\begin{acknowledgments}
	This work is mainly supported by NSFC (No. 12074096), HSFC (No. A2021208013). WY acknowledges support by National Key R$\&$D Program of China (2020YFA0309600). J.L. acknowledges support by the National Natural Science Foundation of China (No. 12274305). We thank Dr. DW Zhai for many helpful discussions. 
\end{acknowledgments}

\appendix

\section{HAMILTONIAN FROM TWO-DIMENSIONAL TO ONE-DIMENSIONAL}

We first discuss the effective Hamiltonian governing the valence band of the K-valley can be expressed as 

\begin{equation}\label{eqiA1}
	H_v=-\frac{p^2}{2m}+\mathcal{U}_v=-\frac{p^2}{2m}+\left(\begin{array}{cccc}
		V_t&H_{op}\\
		H^*_{op}&V_b\\
	\end{array}\right) ,
\end{equation}

\noindent where $-\frac{p^2}{2m}$ describes the quadratic dispersion near the valence band edge, $\mathcal{U}_v$ characterizes the moir{\'e} potential of the valence band coupling, $V_{t/b}$ represents the intralayer moir{\'e} potential, elucidating the band edge shifts in each individual layer due to the charge transfer between layers, and $t/b$ denotes the top/bottom layer \cite{x29}. 

\begin{figure}
	\includegraphics[width=8.5cm]{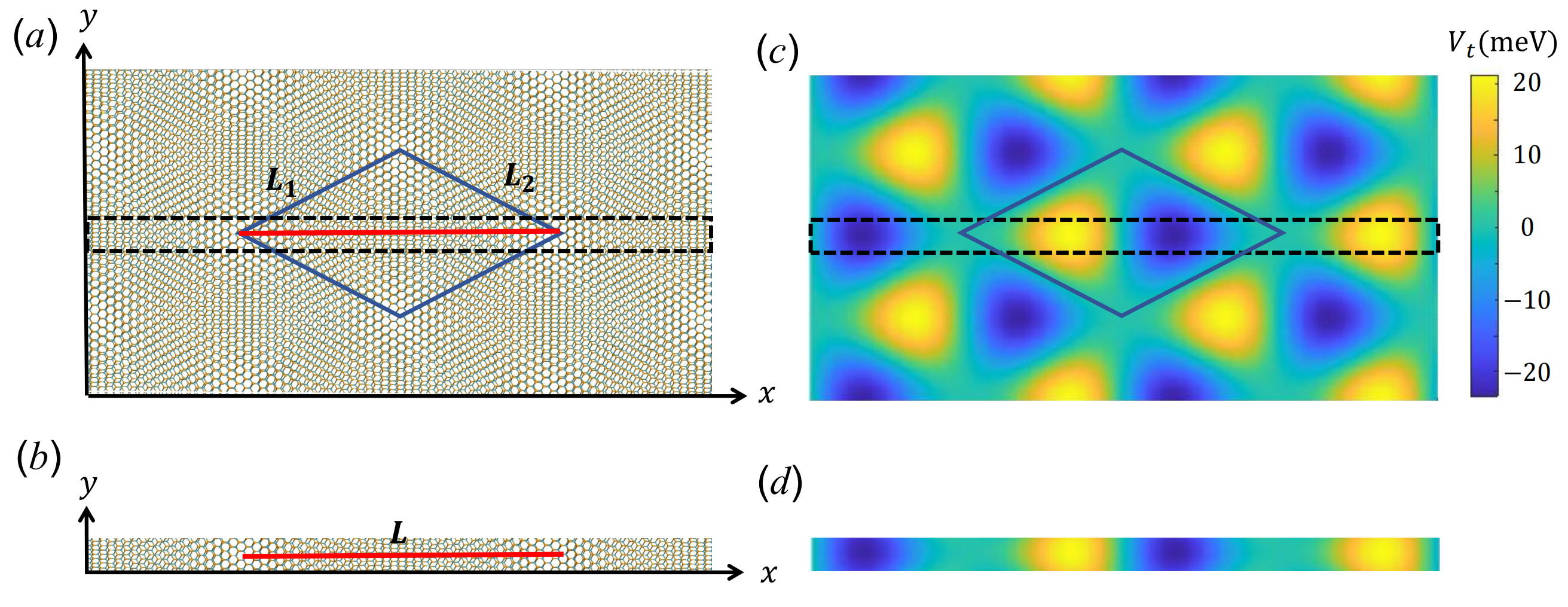}
	\caption{\label{fig:a5}The schematics and potentials of strained moir{\'e} superlattice. (a) The 2D and (b) the 1D strained moir{\'e} superlattice. The top layer moir{\'e} potential $V_t$ corresponding to (c) 2D and (d) 1D.
	}
\end{figure}

We consider the two-dimensional (2D) moir{\'e} superlattice formed in a homobilayer by applying strain to the top layer, as shown in Fig.~\ref{fig:a5}(a), where a blue parallelogram encloses one moiré unit cell with $\boldsymbol{L_i}$ being the moiré primitive lattice vectors. The moiré potential in the 2D case can be modeled as

\begin{equation}\label{eqiA2}
	V_{t/b}=V_0\sum_{i=1}^{3}\cos(\boldsymbol{G_i}\cdot\boldsymbol{r}+\alpha_l) ,
\end{equation}

\begin{equation}\label{eqiA3}
	H_{op}=(\sum_{i=1}^{3}h_0e^{i\boldsymbol{K_i}\cdot\boldsymbol{\delta}}+h_1e^{-i2\boldsymbol{K_i}\cdot\boldsymbol{\delta}})e^{-i\frac{\boldsymbol{A_{\epsilon}}\cdot\boldsymbol{r}}{\hbar}},
\end{equation}

\noindent where $\boldsymbol{G_i} $ denote the three moiré primitive reciprocal lattice vectors, $\boldsymbol{K_i}$ represent the three equivalent monolayer $K$ points, $\boldsymbol{\delta}$ is the local interlayer registry due to strain \cite{x29}. 

In the following, a simpler case of one-dimensional (1D) moir{\'e} is considered. The moir{\'e} potential in the 1D case can be represented by the moir{\'e} potential on the diagonal of the 2D moir{\'e} unit cell (as indicated by the red line in Fig.~\ref{fig:a5} ), i.e., $\boldsymbol{r}=c(\boldsymbol{L_1}+\boldsymbol{L_2})$, where $c\in[0,1]$, which can also be expressed as $c=\frac{x}{L}$. By cutting the 2D structure and extracting the diagonal region of the 2D moir{\'e} unit cell, as indicated by the black dashed line in Fig.~\ref{fig:a5}(a), we effectively obtained a 1D structure, as shown in Fig.~\ref{fig:a5}(b). Fig.~\ref{fig:a5}(c) depicts the top layer moir{\'e} potential $V_t$ in the case of a 2D superlattice, where the moir{\'e} potential corresponding to the 1D case represented by the black dashed line is extracted, as shown in Fig.~\ref{fig:a5}(d). It is observed that the moir{\'e} potential depicted in Fig.~\ref{fig:a5}(d) exhibits only minor differences compared to the 1D moir{\'e} potential represented by our formulated equation, stemming from the neglected lattice mismatch along the $\hat{y}$ direction.

Monolayer primitive vectors are chosen as $\boldsymbol{a_1}=(a,0)$ and $\boldsymbol{a_2}=(\frac{a}{2},\frac{\sqrt{3}a}{2})$. The corresponding monolayer reciprocal lattice vectors are $\boldsymbol{b_1}=(\frac{2\pi}{a},-\frac{2\pi}{\sqrt{3}a})$, $\boldsymbol{b_2}=(0,\frac{4\pi}{\sqrt{3}a})$ and $\boldsymbol{b_3}=-\boldsymbol{b_1}-\boldsymbol{b_2}=(-\frac{2\pi}{a},-\frac{2\pi}{\sqrt{3}a}) $. The displacement vector, moir{\'e} primitive vectors, and reciprocal lattice vectors are then given by $\boldsymbol{\delta}=(I-S^{-1})\boldsymbol{r}$, $\boldsymbol{L_i}=(I-S^{-1})^{-1}\boldsymbol{a_i}$, $\boldsymbol{G_i}=(I-S^{-1})\boldsymbol{b_i}$ respectively \cite{x42}, where $I$ is the identity matrix and $S=\begin{pmatrix}
	1+\epsilon&0\\
	0&1-\epsilon
\end{pmatrix}$ is the strain matrix with strain tensor $\epsilon$. The three equivalent monolayer $K$ points are chosen as $\boldsymbol{K_1}=\frac{2\boldsymbol{b_1}+\boldsymbol{b_2}}{3}=(\frac{4\pi}{3a},0)$, $\boldsymbol{K_2}=(-\frac{2\pi}{3a},\frac{2\pi}{\sqrt{3}a})$ and $\boldsymbol{K_3}=(-\frac{2\pi}{3a},-\frac{2\pi}{\sqrt{3}a})$ \cite{x29}. By substituting the above-mentioned data into Eqs. (\ref{eqiA1}), (\ref{eqiA2}) and (\ref{eqiA3}), we obtain the 1D moir{\'e} potential:

\begin{align}\label{eqiA4}
	V_{t/b}=&V_0\sum_{i=1}^{3}\cos(\boldsymbol{G_i}\cdot\boldsymbol{r}+\alpha_l)\notag\\
	=&V_0[\cos(\boldsymbol{G_1}\cdot\boldsymbol{r}+\alpha_l)+\cos(\boldsymbol{G_2}\cdot\boldsymbol{r}+\alpha_l)\notag\\
	&+\cos(\boldsymbol{G_3}\cdot\boldsymbol{r}+\alpha_l)]\notag\\
	=&V_0[\cos(\frac{2\pi x}{L}+\alpha_l)+\cos(\frac{2\pi x}{L}+\alpha_l)\notag\\
	&+\cos(\frac{-4\pi x}{L}+\alpha_l)]\notag\\
	=&V_0[2\cos(\frac{2\pi x}{L}+\alpha_l)+\cos(-\frac{4\pi x}{L}+\alpha_l)] ,
\end{align}

\begin{align}\label{eqiA5}
	H_{op}=&(\sum_{i=1}^{3}h_0e^{i\boldsymbol{K_i}\cdot\boldsymbol{\delta}}+h_1e^{-i2\boldsymbol{K_i}\cdot\boldsymbol{\delta}})e^{-i\frac{\boldsymbol{A_{\epsilon}}\cdot\boldsymbol{r}}{\hbar}}\notag\\
	=&(h_0e^{i\boldsymbol{K_1}\cdot\boldsymbol{\delta}}+h_1e^{-i2\boldsymbol{K_1}\cdot\boldsymbol{\delta}}+h_0e^{i\boldsymbol{K_2}\cdot\boldsymbol{\delta}}+h_1e^{-i2\boldsymbol{K_2}\cdot\boldsymbol{\delta}}\notag\\
	&+h_0e^{i\boldsymbol{K_3}\cdot\boldsymbol{\delta}}+h_1e^{-i2\boldsymbol{K_3}\cdot\boldsymbol{\delta}})e^{-i\frac{\boldsymbol{A_{\epsilon}}\cdot\boldsymbol{r}}{\hbar}}\notag\\
	=&[h_0e^{i\frac{2\pi x}{L}}+h_1e^{-i2\frac{2\pi x}{L}}+h_0+h_1\notag\\
	&+h_0e^{i(-\frac{2\pi x}{L})}+h_1e^{-i2(-\frac{2\pi x}{L})}]e^{-i\frac{\boldsymbol{A_{\epsilon}}\cdot\boldsymbol{r}}{\hbar}}\notag\\
	=&[h_0(1+e^{i\frac{2\pi x}{L}}+e^{-i\frac{2\pi x}{L}})\notag\\
	&+h_1(1+e^{-i\frac{4\pi x}{L}}+e^{i\frac{4\pi x}{L}})]e^{-i\frac{\boldsymbol{A_{\epsilon}}\cdot\boldsymbol{r}}{\hbar}}\notag\\
	=&\{h_0[1+\cos(\frac{2\pi x}{L})+i\sin(\frac{2\pi x}{L})+\cos(\frac{2\pi x}{L})\notag\\
	&-i\sin(\frac{2\pi x}{L})]+h_1[1+\cos(\frac{4\pi x}{L})-i\sin(\frac{4\pi x}{L})\notag\\
	&+\cos(\frac{4\pi x}{L})+i\sin(\frac{4\pi x}{L})]\}e^{-i\frac{\boldsymbol{A_{\epsilon}}\cdot\boldsymbol{r}}{\hbar}}\notag\\
	=&\{h_0[1+2\cos(\frac{2\pi x}{L})]+h_1[1+2\cos(\frac{4\pi x}{L})]\}e^{-i\frac{\boldsymbol{A_{\epsilon}}\cdot\boldsymbol{r}}{\hbar}}.
\end{align}

\section{DISCRETE LATTICE MODEL}

\begin{figure}
	\includegraphics[width=8.5cm]{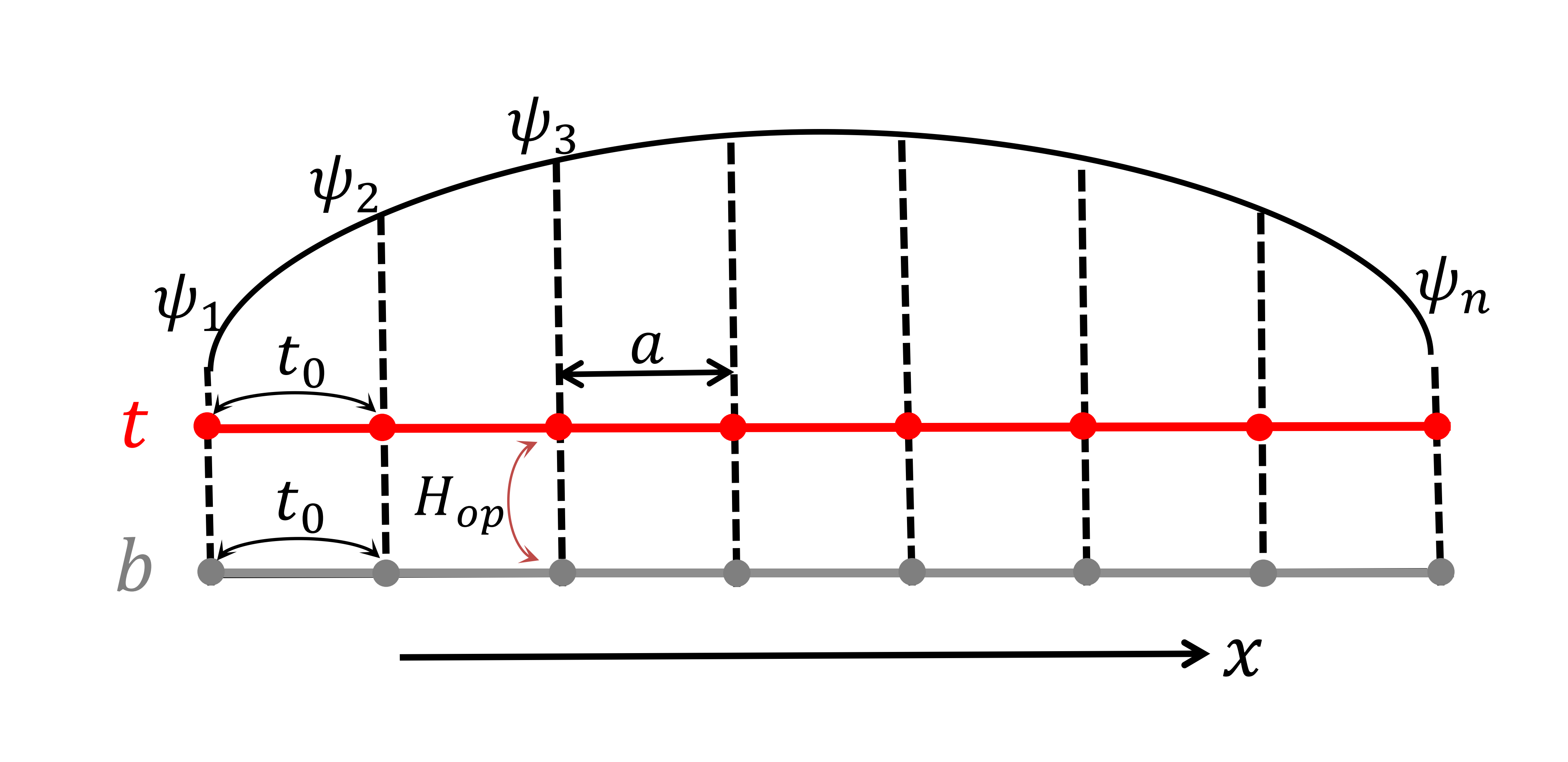}
	\caption{\label{fig:a6}A continuous function can be represented by its values at a set of points on a discrete lattice. 
	}
\end{figure}

In this paper, we employ the finite difference method to solve for the Hamiltonian of the homobilayer TMDs. The basic idea of the finite difference method is to replace continuous variable differential equations with discrete difference equations involving a finite number of unknowns. This conversion can be done in many ways, but the simplest one is to choose a discrete lattice. To see how this is done let us for simplicity consider just one dimension and discretize the position variable $x$ into a lattice as shown in Fig.~\ref{fig:a6}: $x_n=na$.

The next step is to obtain the matrix representing the Hamiltonian operator

\begin{align}\label{eqiB1}
	H_v=&-\frac{p^2}{2m}+\mathcal{U}_v\notag\\
	=&\begin{pmatrix}
		V_t+\frac{\hbar^2}{2m}\frac{\partial^2}{\partial x^2}&H_{op}\\
		H_{op}^*&	V_b+\frac{\hbar^2}{2m}\frac{\partial^2}{\partial x^2}
	\end{pmatrix} \notag\\
	=&\begin{pmatrix}
		H_t&H_{tb}\\
		H_{bt}&H_b	
	\end{pmatrix} ,
\end{align}

\noindent where $H_t$ and $H_b$ represent the Hamiltonians of the top and bottom layers, respectively. $H_{tb}$ and $H_{bt}$ are the Hamiltonians describing the coupling between the top and bottom layers. The conversion of differential equations into difference equations necessitates the application of the finite difference method:

\begin{equation}\label{eqiB2}
	(\frac{\partial^2 \Psi}{\partial x^2})_{x=x_n}\to\frac{1}{a^2}(\Psi_{n+1}-2\Psi_n+\Psi_{n-1}) .
\end{equation}

This allows us to write: 

\begin{align}\label{eqiB3}
	[H_t\psi]_{x=x_n} =&V_{tn}\psi_n+\frac{\hbar^2}{2ma^2}[\psi_{n+1}-2\psi_n+\psi_{n-1}]\notag\\
	=&(V_{tn}-2t_0)\psi_n+t_0\psi_{n+1}+t_0\psi_{n-1}\notag\\
	=&(V_{tn}-2t_0)\delta_{n,m}+t_0\delta_{n,m+1}+t_0\delta_{n,m-1},
\end{align}

and

\begin{align}\label{eqiB4}
	[H_b\psi]_{x=x_n} =&V_{bn}\psi_n+\frac{\hbar^2}{2ma^2}[\psi_{n+1}-2\psi_n+\psi_{n-1}]\notag\\
	=&(V_{bn}-2t_0)\psi_n+t_0\psi_{n+1}+t_0\psi_{n-1}\notag\\
	=&(V_{bn}-2t_0)\delta_{n,m}+t_0\delta_{n,m+1}+t_0\delta_{n,m-1},
\end{align}

\noindent where $t_0\equiv\frac{\hbar^2}{2ma^2}$, $V_{tn}\equiv V_t(x_n)$, $V_{bn}\equiv V_b(x_n)$ and $\delta_{n,m}$ is the Kronecker delta. This means that matrix representing $H_{l=t/b}$, $H_{bt}$ and $H_{tb}$ can be expressed as

\[H_t=\begin{pmatrix}
	V_{t1}-2t_0&t_0&\dots&0&0\\
	t_0&V_{t2}-2t_0&\dots&0&0\\
	\vdots&\vdots&\ddots&\vdots&\vdots\\
	0&0&\dots&V_{t(N-1)}-2t_0&t_0\\
	0&0&\dots&t_0&V_{tN}-2t_0
\end{pmatrix}\]

\[H_b=\begin{pmatrix}
	V_{b1}-2t_0&t_0&\dots&0&0\\
	t_0&V_{b2}-2t_0&\dots&0&0\\
	\vdots&\vdots&\ddots&\vdots&\vdots\\
	0&0&\dots&V_{b(N-1)}-2t_0&t_0\\
	0&0&\dots&t_0&V_{bN}-2t_0
\end{pmatrix}\]

\[H_{tb}=\begin{pmatrix}
	H_{op1}&0&\dots&0&0\\
	0&H_{op2}&\dots&0&0\\
	\vdots&\vdots&\ddots&\vdots&\vdots\\
	0&0&\dots&H_{op(N-1)}&0\\
	0&0&\dots&0&H_{opN}
\end{pmatrix}\]

\[H_{bt}=\begin{pmatrix}
	H^*_{op1}&0&\dots&0&0\\
	0&H^*_{op2}&\dots&0&0\\
	\vdots&\vdots&\ddots&\vdots&\vdots\\
	0&0&\dots&H^*_{op(N-1)}&0\\
	0&0&\dots&0&H^*_{opN}
\end{pmatrix}\]

\section{RECURSIVE GREEN’S FUNCTION TECHNIQUE }

Based on the one-dimensional model studied in this paper, we consider semi-infinite quantum wires as source and drain. Labeling each sites with an index $i$, the equation of motion can be written as \cite{x39}:

\begin{equation}\label{eqiC1}
	-(E\boldsymbol{I}-\boldsymbol{H}_i)\boldsymbol{c}_i+\boldsymbol{H_{i,i-1}}\boldsymbol{c_{i-1}}+\boldsymbol{H_{i,i+1}}\boldsymbol{c_{i+1}}=0 ,
\end{equation}

\noindent where $\boldsymbol{c_i}$ is a vector of dimension 2 describing the top-layer and bottom-layer wave-function coefficients on site $i$ within discrete lattice model, the $2\times2$ matrix $\boldsymbol{H_{i,i-1}}$ ($\boldsymbol{H_{i,i+1}}$) describes the interaction between lattice site $i$ and lattice site $i-1$ ($i+1$).

The equation of motion can be rewritten as a transfer matrix form

\begin{equation}\label{eqiC2}
	\begin{aligned}
	\begin{pmatrix}
			\boldsymbol{c}_{i+1} \\
			\boldsymbol{c}_{i}
		\end{pmatrix}=
		\begin{pmatrix}
			\boldsymbol{H}^{-1}_{i, i+1}(E \boldsymbol{I}-\boldsymbol{H}_{i}) & -\boldsymbol{H}^{-1}_{i, i+1} \boldsymbol{H}_{i, i-1} \\
			\boldsymbol{I} & 0 
		\end{pmatrix} 
		\begin{pmatrix}
			\boldsymbol{c}_{i} \\
			\boldsymbol{c}_{i-1}
		\end{pmatrix}.
	\end{aligned}
\end{equation}

\noindent We suppose the solutions of Eq. (\ref{eqiC1}) have Bloch symmetry, $\boldsymbol{c}_i=\lambda\boldsymbol{c}_{i-1}$ and $\boldsymbol{c}_{i+1}=\lambda^2\boldsymbol{c}_{i-1}$. Substituting this into Eq. (\ref{eqiC2}) results in an eigenvalue problem,

\begin{equation}\label{eqiC3}
	\begin{aligned}
		\begin{pmatrix}
			\boldsymbol{H}^{-1}_{i, i+1}(E \boldsymbol{I}-\boldsymbol{H}_{i}) & -\boldsymbol{H}^{-1}_{i, i+1} \boldsymbol{H}_{i, i-1} \\
			\boldsymbol{I} & 0
		\end{pmatrix}
		\begin{pmatrix}
			\boldsymbol{c}_{i} \\
			\boldsymbol{c}_{i-1}
		\end{pmatrix}=\lambda
		\begin{pmatrix}
			\boldsymbol{c}_{i} \\
			\boldsymbol{c}_{i-1} 
		\end{pmatrix} .
	\end{aligned}
\end{equation}

\noindent This equation has $2\times2$ solutions, which are classified into two right-going modes and two left-going modes. The right-going modes consist of evanescent waves that are decaying to the right and traveling waves that are propagating to the right. Similarly, left-going modes are decaying or propagating to the left. One can distinguish right- from left-going modes on the basis of the sign of the Bloch velocity. The propagating modes have $\left|\lambda\right|=1$ and the evanescent modes have $\left|\lambda\right|\neq1$. The eigenvalue $\lambda$ is related to the wave number $k$ through $\lambda=\exp(ika)$. The eigenvalues are denoted by $\lambda_n(\pm)$ where $n=1,2$, the corresponding eigenvectors by $\boldsymbol{u}_n(\pm)$, where the right-going and left-going modes are labeled as $(+)$ and $(-)$. Let $\boldsymbol{u}_1(-)$, $\boldsymbol{u}_2(-)$ be $\boldsymbol{c}_0$ of the left-going solutions corresponding to $\lambda_1(-)$, $\lambda_2(-)$ and $\boldsymbol{u}_1(+)$, $\boldsymbol{u}_2(+)$, be $\boldsymbol{c}_0$ of the right-going solutions corresponding to $\lambda_1(+)$, $\lambda_2(+)$. The Bloch velocities are given by the expression

\begin{equation}\label{eqiC4}
	v_{n}(\pm)=-\frac{2 a}{\hbar} \operatorname{Im}\left[\lambda_{n}(\pm) \boldsymbol{u}_{n}^{\dagger}(\pm)\boldsymbol{H}_{i, i+1}^{\dagger} \boldsymbol{u}_{n}(\pm)\right].
\end{equation}

\noindent Define 

\begin{equation}\label{eqiC5}
	U({\pm})=\left(\boldsymbol{u}_1(\pm) \boldsymbol{u}_2(\pm)\right),
\end{equation}

\noindent and

\begin{equation}\label{eqiC6}
	\Lambda(\pm)=\begin{pmatrix}
		\lambda_{1}(\pm) &  \\
		 & \lambda_{2}(\pm)
	\end{pmatrix}.
\end{equation}

\noindent Any left- and right-going waves are written, at $i=0$, for example, as

\begin{equation}\label{eqiC7}
	\boldsymbol{c}_{i}({\pm})=U(\pm) \boldsymbol{c}(\pm),
\end{equation}

\noindent where $\boldsymbol{c}(\pm)$ is an appropriate vector consisting of expansion coefficients. For general $i$ we have

\begin{equation}\label{eqiC8}
	\boldsymbol{c}_{i}({\pm})=U(\pm)\Lambda^{i}(\pm) \boldsymbol{c}(\pm),
\end{equation}

\noindent which leads to the relation

\begin{equation}\label{eqiC9}
	\boldsymbol{c}_{i}({\pm})=F^{i-i'}(\pm)\boldsymbol{c}_{i'}(\pm),
\end{equation}

\noindent with

\begin{equation}\label{eqiC10}
	F({\pm})=U({\pm}) \Lambda({\pm}) U^{-1}({\pm}).
\end{equation}

Next, the scattering problem in the quantum wire with a length of $N_x$ is considered. We separate the amplitude $\boldsymbol{c}_0$ at site $0$ into the right-going and left-going solutions, 

\begin{equation}\label{eqiC11}
	\boldsymbol{c}_0=\boldsymbol{c}_0(+)+\boldsymbol{c}_0(-).
\end{equation}

\noindent Using the equation of motion at site $0$ 

\begin{equation}\label{eqiC12}
	-(E\boldsymbol{I}-\boldsymbol{H}_0)\boldsymbol{c}_0+\boldsymbol{H}_{0,-1}\boldsymbol{c}_{-1}+\boldsymbol{H}_{0,1}\boldsymbol{c}_1=0 ,
\end{equation}

\noindent and the amplitude at site $-1$

\begin{equation}\label{eqiC13}
	\boldsymbol{c}_{-1}=F^{-1}(-)\boldsymbol{c}_0+[F^{-1}(+)-F^{-1}(-)]\boldsymbol{c}_0(+) ,
\end{equation}

\noindent we have 

\begin{equation}\label{eqiC14}
	(E\boldsymbol{I}-\tilde{\boldsymbol{H}}_0)\boldsymbol{c}_0-\boldsymbol{H}_{0,1}\boldsymbol{c}_{1}=\boldsymbol{H}_{0,-1}[F^{-1}(+)-F^{-1}(-)]\boldsymbol{c}_0(+) ,
\end{equation}

\noindent where $\tilde{\boldsymbol{H}}_0=\boldsymbol{H}_{0}+\boldsymbol{H}_{0,-1}F^{-1}(-)$.

On the other hand, there is no reflection (left-going waves) at site $N_{x}+1$, that is

\begin{equation}\label{eqiC15}
	\boldsymbol{c}_{N_{x}+2}=F(+)\boldsymbol{c}_{N_{x}+1}.
\end{equation}

\noindent Therefore, we have 

\begin{equation}\label{eqiC16}
	(E\boldsymbol{I}-\tilde{\boldsymbol{H}}_{N_{x}+1})\boldsymbol{c}_{N_{x}+1}-\boldsymbol{H}_{N_{x}+1,N_{x}}\boldsymbol{c}_{N_{x}}=0 ,
\end{equation}

\noindent where $\tilde{\boldsymbol{H}}_{N_{x}+1}=\boldsymbol{H}_{N_{x}+1}+\boldsymbol{H}_{N_{x}+1,N_{x}+2}F(+) $.

Define the Green function as

\begin{equation}\label{eqiC17}
	\boldsymbol{G}=\frac{1}{[E\boldsymbol{I}-\tilde{\boldsymbol{H}}]} ,
\end{equation}

\noindent where 

\begin{equation}\label{eqiC18}
	\tilde{\boldsymbol{H}}=\begin{pmatrix}
		\tilde{\boldsymbol{H}}_0&\boldsymbol{H}_{0,1}&0&\dots&0&0\\
		\boldsymbol{H}_{1,0}&\tilde{\boldsymbol{H}}_1&\boldsymbol{H}_{1,2}&\dots&0&0\\
		0&\boldsymbol{H}_{2,1}&\tilde{\boldsymbol{H}}_2&\dots&0&0\\
		\vdots&\vdots&\vdots&\ddots&\vdots&\vdots\\
		0&0&0&\dots&\tilde{\boldsymbol{H}}_{N_x}&\boldsymbol{H}_{N_{x},N_{x}+1}\\
		0&0&0&\dots&\boldsymbol{H}_{N_{x}+1,N_{x}}&\tilde{\boldsymbol{H}}_{N_{x}+1}
	\end{pmatrix},
\end{equation}

\noindent with $\tilde{\boldsymbol{H}}_{i}=\boldsymbol{H}_{i}$ for $i=1,2,\dots,N_x$. We can derive

\begin{align}\label{eqiC19}
		\boldsymbol{c}_{N_{x}+1}(+)=&\boldsymbol{c}_{N_{x}+1}\notag\\
		=&-\boldsymbol{G}_{N_{x}+1,0}\boldsymbol{H}_{0,-1}[F^{-1}(+)-F^{-1}(-)]\boldsymbol{c}_{0}(+),
\end{align}

\noindent and

\begin{align}\label{eqiC20}
		\boldsymbol{c}_{0}(-)=&\boldsymbol{c}_{0}-\boldsymbol{c}_{0}(+)\notag\\
		=&\{-\boldsymbol{G}_{0,0}\boldsymbol{H}_{0,-1}[F^{-1}(+)-F^{-1}(-)]-\boldsymbol{I}\}\boldsymbol{c}_{0}(+),
\end{align}

\noindent The transmission coefficient for the incident layer $m$ ($m=1,2$) with velocity $v_m$ and out-going layer $n$ ($n=1,2$) with velocity $v_n$ can be obtained as:

\begin{align}\label{eqiC21}
		t_{m n}=&\sqrt{\frac{v_{n}}{v_{m}}}\{ -U^{-1}(+)\boldsymbol{G}_{N_{x}+1,0}\boldsymbol{H}_{0,-1}\notag\\
		&[F^{-1}(+)-F^{-1}(-)] U(+)\}_{m n},
\end{align}

\noindent and the reflection coefficient for the incident layer $m$ and out-going layer $n$ as

\begin{align}\label{eqiC22}
		r_{m n}=&\sqrt{\frac{v_{n}}{v_{m}}} ( U^{-1}(-)\{-\boldsymbol{G}_{0,0}\boldsymbol{H}_{0,-1}\notag\\
		&[F^{-1}(+)-F^{-1}(-)] -\boldsymbol{I}\}U(+))_{m n}.
\end{align}

\noindent The Green’s-function matrix block $\boldsymbol{G}_{N_{x}+1,0}$ and $\boldsymbol{G}_{0,0}$ can be found using a set of 
recursive formulas,

\begin{equation}\label{eqiC23}
	\boldsymbol{G}_{i+1,i+1}=(E\boldsymbol{I}-\tilde{\boldsymbol{H}}_{i+1}-\boldsymbol{H}_{i+1,i}\boldsymbol{G}_{i,i}\boldsymbol{H}_{i,i+1})^{-1},
\end{equation}

\begin{equation}\label{eqiC24}
	\boldsymbol{G}_{i+1,0}=\boldsymbol{G}_{i+1,i+1}\boldsymbol{H}_{i+1,i}\boldsymbol{G}_{i,0},
\end{equation}

\noindent with the initial conditions $\boldsymbol{G}_{0,0}=[E\boldsymbol{I}-\tilde{\boldsymbol{H}}_{0}]^{-1} $ and $\boldsymbol{G}_{N_{x}+1,N_{x}+1}=[E\boldsymbol{I}-\tilde{\boldsymbol{H}}_{N_{x}+1}]^{-1} $, respectively.

\bibliography{refernces.bib}

\end{document}